\documentclass{aa}  

\usepackage{graphicx}
\usepackage{txfonts}
\begin{document}

   \title{Gaia EDR3 distances of the young stellar clusters in the extended Carina Nebula complex\thanks{Table 1 
is only available in electronic form
at the CDS via anonymous ftp to cdsarc.u-strasbg.fr (130.79.128.5)
or via http://cdsweb.u-strasbg.fr/cgi-bin/qcat?J/A+A/}}

   \author{C. G{\"o}ppl
          \and
          T. Preibisch\thanks{corresponding author}
          }

   \institute{Universit\"ats-Sternwarte M\"unchen, 
              Ludwig-Maximilians-Universit\"at,
              Scheinerstr.~1, 81679 M\"unchen, Germany\\
              \email{preibisch@usm.uni-muenchen.de}
             }

\titlerunning{Gaia distances to the Carina Nebula Complex}
\authorrunning{G{\"o}ppl \& Preibisch}

   \date{Received 3 November 2021; accepted 14 January 2022}

 
  \abstract
   {The Carina Nebula complex (CNC) is one of the most massive and active star-forming 
regions in our Galaxy and it contains several large young star clusters. 
The distances
  of the individual clusters and their physical connection were poorly known up to now,
 with strongly discrepant results reported in the literature.}
   {We want to determine reliable distances
    of the young stellar clusters in the central Carina Nebula
    region (in particular, Tr~14, 15, and 16) and the prominent clusters NGC~3324 and NGC~3293
    in the northwestern periphery of the CNC.
   }
   {We analyzed the parallaxes in Gaia EDR3 for a comprehensive sample
    of 237 spectroscopically identified OB stars, 
   as well as for 9562 X-ray-selected young stars throughout the complex.
 We also performed an astrometric analysis to identify members of the young cluster vdBH~99, 
   which is located in the foreground of the northwestern part of the Carina Nebula.
   }
   {We find that the distances of the investigated clusters in the CNC are equal within
 $\le 2\%$, and yield very consistent most likely mean distance values of 
 \mbox{$ 2.36^{+0.05}_{-0.05}$~kpc} for the OB star sample
 and \mbox{$2.34^{+0.05}_{-0.06}$~kpc} for the sample of X-ray-selected young stars. 
   }
   {Our results show that the clusters in the CNC constitute
    a coherent star-forming region, in particular with regard to NGC~3324 and NGC~3293 at the
    northwestern periphery, which are (within $\le 2\%$) at the same distance as the 
  central Carina Nebula.
  For the foreground cluster vdBH~99, we find a mean distance of 
$441^{+2}_{-2}$~pc  and an age of $\simeq 60$~Myr. We quantified the contamination of X-ray-selected samples of Carina Nebula stars
  based on members of this foreground cluster.}


   \maketitle
%

\section{Introduction}

\begin{figure}  
\includegraphics[width=8.85cm,trim=50 175 55 175,clip]{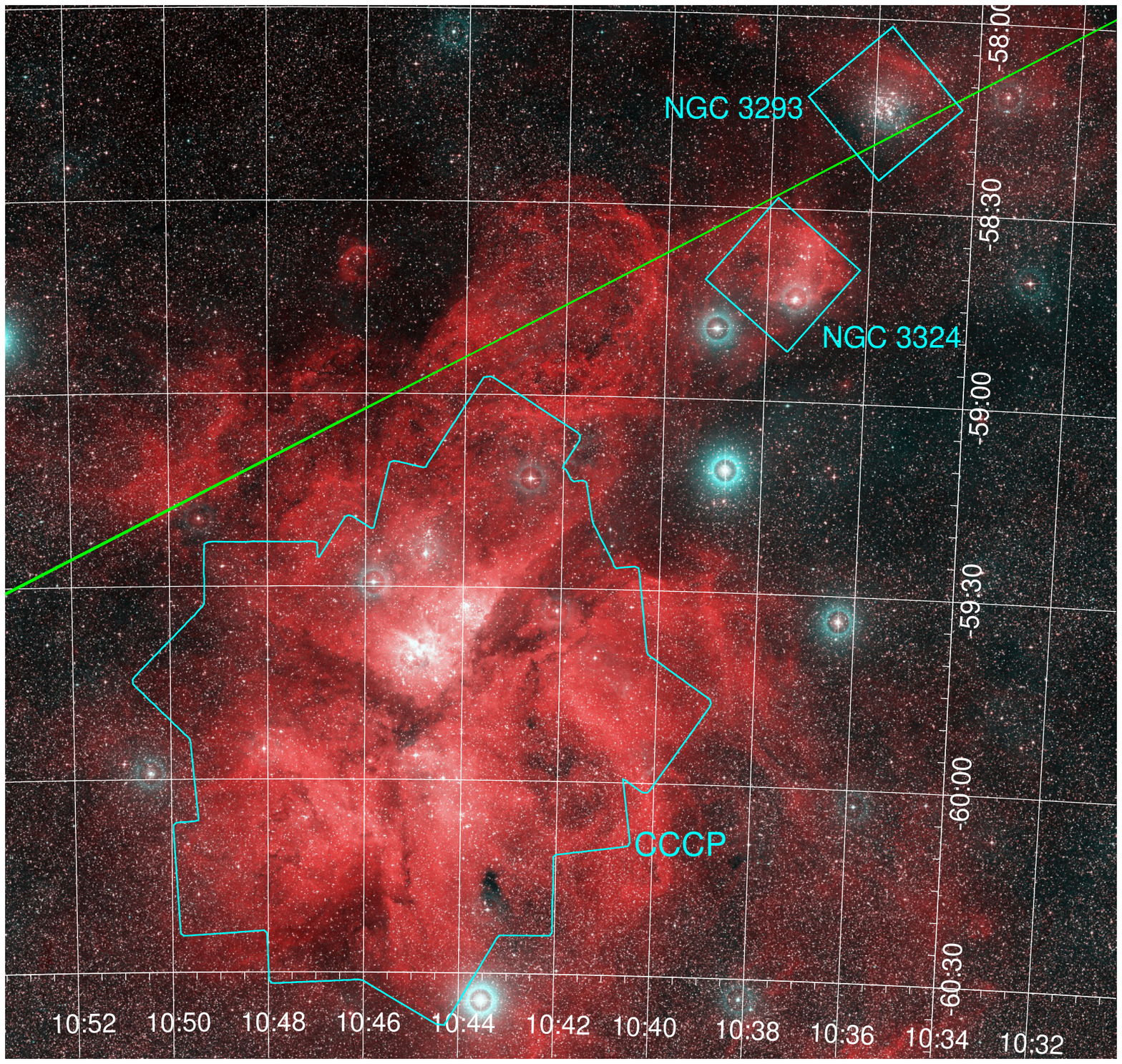}
\caption{Optical image of the Carina Nebula complex 
from the ESO Photo Release eso0905b 
(image credit: ESO/Digitized Sky Survey 2, Davide De Martin).
North is up and east to the left. The white grid shows the J2000
celestial coordinates and the green line marks the galactic plane.
The X-ray survey region of the \textit{Chandra} Carina Complex
Project (CCCP) as well as the additional \textit{Chandra} observations of NGC~3324
and NGC~3293 are marked by the cyan outlines and labeled. The clusters Tr~14, 15, 
and 16 are located in the central bright part of the nebulosity.
\label{CNC.fig} }
\end{figure}

The Carina Nebula \citep[NGC 3372; see][for a review and Fig.~\ref{CNC.fig} for an optical image]{SB08} is
a very prominent giant \ion{H}{II} region that is powered by
at least 70 O-type and Wolf-Rayet (WR) stars, including many of the
most luminous and massive stars known in our Galaxy \citep[see][]{Smith06}.
The central $\sim 20\arcmin$  region of the Carina Nebula is
dominated by the young star clusters Tr~14, 15, and 16,
along with several smaller clusters  such as Bochum~10 and 11 or the Treasure Chest cluster
are found south  of Tr~14 -- 16.
However, the optical nebulosity of the Carina Nebula 
\citep[see][]{2000ApJ...532L.145S}, the dusty clouds
\citep[see][]{Preibisch12}, and the molecular gas \citep[see][]{2016MNRAS.456.2406R}
extend over a much wider area spanning more than two degrees on the sky (see Fig.~\ref{CNC.fig}).
This large cloud complex contains a total gas and dust mass of about $10^6\,M_\odot$
\citep{Preibisch12} and harbors several further star clusters.
The clouds south of the central Carina Nebula 
constitute the so-called ``Southern Pillars'' region 
\citep{Smith10b}, where the ionizing radiation from the massive stars in Tr~14 and 16
is currently triggering the formation of a new generation of stars in the 
remaining dense clouds.
The area northwest of the central Carina Nebula contains the prominent
\ion{H}{ii} region Gum~31 around the stellar cluster NGC~3324
and, just beyond the northwestern edge of the optical nebulosity,
the young stellar cluster NGC~3293.
We denote this extended ($\approx 2\degr$) complex of star clusters, optical nebulosity,
and dusty clouds as the Carina Nebula complex (CNC) in the following text.

Comprehensive information about the young stellar
populations in the Carina Nebula was obtained thanks to the
deep X-ray imaging survey of the
{\it Chandra} Carina Complex Project \citep[CCCP; see][for an overview]{CCCP-intro},
which mapped the central 1.4 square-degrees 
(i.e.~about half of the total spatial extent of the CNC)
with the \textit{Chandra} X-ray observatory and revealed
14\,368 individual X-ray sources.
The combination of the X-ray data with deep infrared
surveys \citep{Povich11,HAWKI-survey,VISTA1} showed that 
the large majority of these X-ray sources are young stars in the Carina Nebula
\citep[see][]{2011ApJS..194....4B},
with ages ranging from $\la 1$~Myr up to $\approx 10$~Myr \citep{CCCP-HAWKI,2011ApJS..194...11W,2011ApJS..194...12W}.
The northern clusters  NGC~3324 and  NGC~3293 have been studied in a similar way
by individual \textit{Chandra} X-ray observations \citep[see][]{Preibisch14,2017A&A...605A..85P},
which revealed their young stellar populations by correlation with infrared data.

Given the large angular extent of the CNC on the sky, it was not clear up to now
whether the individual clusters of the CNC are physically related.
Since the region is very close to the galactic plane,
the possibility of chance projections with stellar clusters in the galactic background
must not be neglected.
In particular, the physical relation between the clusters 
NGC~3324 and NGC~3293 at the northwestern periphery of the CNC and the clusters
Tr~14 -- 16 in the
central parts of the Carina Nebula remained dubious, since widely discrepant
distance estimates had been reported in the literature.

\medskip

The pre-Gaia
distance estimates to stellar clusters in the Carina Nebula have often yielded 
conflicting results and also caused high uncertainties
in terms of the inferred stellar parameters, such as ages and masses, as well
as the question of whether parts of the extended complex are actually physically related
or whether they only appear to be in proximity on the sky due to projection effects.
The distance to $\eta$~Car in the center of the Carina Nebula
 could be measured with good accuracy
by the kinematical analysis of the expanding Homunculus Nebula, yielding values of
$(2.2 \pm 0.2)$~kpc \citep{1993PASAu..10..338A}
and $(2.35 \pm 0.05)$~kpc \citep{Smith06}. 
For the clusters Tr~14 -- 16, substantially discrepant distance values
had been reported; tabulations of literature results and 
discussions about the reasons for the discrepancies can be found in
\citet[][(see their Section 3.3 and Table~A.1;]{MA2020} and 
\citet[][their Table~1]{2021ApJ...914...18S}. 
Although \citet{Smith06} and \citet{SB08} provided good arguments for a 
common distance of $(2.3 \pm 0.1)$~kpc for 
Tr~14 -- 16, some recent studies have suggested either significantly different distances
for the three clusters \citep[e.g., 
$\approx 2.8$~kpc for Tr~14, $\approx 2.1$~kpc for Tr~15, 
and $\approx 2.8$~kpc 
for Tr~16 obtained by][]{2005A&A...438.1163K},
or claimed considerably larger values 
\citep[e.g.~ $\approx 5.7$~kpc for Tr~14, $\approx 3.4$~kpc for Tr~15, 
and $\approx 5.4$~kpc for Tr~16  according to][based on optical photometry]{2010MNRAS.403.1491P}.

\medskip

For the cluster NGC~3324, located just northwest of the Carina Nebula,
distances ranging from 
$\approx 2.3$~kpc \citep{2005A&A...438.1163K},
over    $\approx 2.6$~kpc \citep{1982AJ.....87.1300W},
$(3.0 \pm 0.1)$~kpc \citep{2001A&A...371..107C},
and $\approx 3.1$~kpc \citep{1977A&AS...27..145C}, 
and up to
$\approx 4.3$~kpc \citep{2010MNRAS.403.1491P} can be found in the literature.
Also for NGC~3293, located at the northwestern periphery of the CNC,
no clear distance information was available
up to now. Reported values include
$\approx 2.3$~kpc \citep{Dias02},
$(2.75 \pm 0.05)$~kpc \citep{2003A&A...402..549B},
$\approx 2.5$~kpc \citep{2005A&A...438.1163K},
$\approx 3.4$~kpc \citep{2010MNRAS.403.1491P}, and
$2.7^{+0.7}_{-1.1}$~kpc  \citep{2012PASP..124..128K}.
Most of these literature values would suggest that NGC~3324 and
NGC~3293 are significantly more distant than the central Carina Nebula,
implying that there is no physical connection but only 
a projection effect.

However, most of the quoted pre-Gaia distance estimates are quite uncertain
since they were
based on the analysis of optical color-magnitude diagrams (CMDs),
which can be strongly affected by the assumptions about the
reddening of the stars in each cluster.
It is well known that the cloud extinction in the CNC shows
very strong spatial variations on small scales; stars with extinctions of 
$A_V \la 2$~mag are found
close to stars suffering extinctions of $A_V \approx 10$~mag or 
more\footnote{A good example for the strong spatial variations of cloud extinction
in the Carina Nebula is
the highly obscured cluster Tr~16-SE \cite[see][]{Tr16-SE-KMOS}: the stars in Tr~16-SE suffer extinctions
of $A_V \ga 5 - 10$~mag, making them very faint or invisible in optical images,
although they are located just a few arcminutes south of the (much less obscured and therefore
optically much brighter) stars of the cluster Tr~16.} \citep[see][for a cloud column-density map]{Preibisch12}.
These strong variations in the extinction of individual stars
 can cause large uncertainties in the cluster distance estimates derived
from CMDs.

\medskip

The Gaia mission \citep{Gaia} is currently providing 
astrometric data with unprecedented precision, and allows 
for direct measurements of distances to be made for the stars in the Carina Nebula for the first time.
The study of \citet{2018A&A...618A..93C} used on Gaia DR2 parallaxes to 
derive mean distances of $2.414^{+0.009}_{-0.009}$~kpc for Tr~14, 
$2.339^{+0.012}_{-0.010}$~kpc for Tr~15, and 
$2.395^{+0.010}_{-0.012}$~kpc for Tr~16, where
the quoted uncertainties are based on the 16th to 84th percentile distance 
confidence interval.
\citet{MA2020} determined very similar distances of
$2.43^{+0.29}_{-0.23}$~kpc for Tr~14
and
$2.38^{+0.27}_{-0.22}$~kpc for 
Tr~16~W.
These values are in good agreement to the previous
``canonical'' distance of $(2.3 \pm 0.1)$~kpc and 
with the expansion-parallax distance of $(2.35 \pm 0.05)$~kpc determined
for the Homunculus Nebula around $\eta$ Car \citep{Smith06}.

The analysis of the Gaia DR2 parallaxes for the
X-ray-selected stars in the Carina Nebula from the CCCP
leads to distance estimates of $2.62_{-0.25}^{+0.31}$~kpc \citep{2019ApJ...870...32K}
and
$2.50^{+0.28}_{-0.23}$~kpc \citep{Povich19};
both values are higher, but still marginally consistent with the previous 
``canonical'' distance.

For NGC~3324, \citet{2018A&A...618A..93C}  determined a distance of
$2.600^{+0.023}_{-0.033}$~kpc from Gaia DR2 data; 
this value would suggest that NGC~3324 
is about 200~pc further away than Tr~14 -- 16, namely,~behind 
the Carina Nebula.
For NGC~3293, \citet{2018A&A...618A..93C} derived a distance of
$2.470^{+0.010}_{-0.009}$~kpc based on Gaia DR2 data.

\bigskip

The recently released Gaia EDR3 data \citep[see][]{Gaia-EDR3}
have considerably improved the accuracy of the astrometric parameters
compared to the DR2 results and clearly a fresh look at the
distances of the stars in the CNC is warranted.
A study of the Gaia EDR3 data for 69 massive stars in
the central Carina Nebula by \cite{2021ApJ...914...18S}
found a common distance of $(2.35 \pm 0.08)$~kpc for these stars
and no significant differences in the distances of the central
Carina Nebula clusters Tr~14, 15, and 16.
As pointed out by the authors,
the controversy over the distance of the stellar clusters in the
central region of the Carina Nebula is now settled thanks to this result.
Very similar distances were determined in the recent study of \citet{MA2021},
who found
$2.363^{+0.061}_{-0.058}$~kpc for Tr~14,
$2.305^{+0.064}_{-0.061}$~kpc for Tr~16~W,
and
$2.311^{+0.058}_{-0.056}$~kpc for Tr~16~E.
These Gaia results confirm that the above mentioned ``canonical'' distance 
value was already a very good estimate.
While the distance to the central Carina Nebula is now well determined,
the distances of the northern clusters NGC~3324 and NGC~3293 in the CNC 
and their relation to the central Carina Nebula stellar population
still remain to be addressed with the new Gaia EDR3 data.

Here we use Gaia EDR3 data to
(1) extend the study of \cite{2021ApJ...914...18S} with a larger 
(237 rather than 69 objects) sample of
OB stars, covering not only the central Carina Nebula but also including 
the clusters NGC~3324 and NGC~3293 at the northwestern periphery of the CNC (Sect.~2);
(2) determine the mean distance to the  large sample of 9562 Gaia matches
to X-ray-selected stars in the CNC and compare it to the distances
derived for the OB stars (Sect.~3);
(3) investigate the stellar cluster vdBH~99 \citep{1975AJ.....80...11V},
which is located in the foreground of the northwestern part
of the CNC, and quantify the contamination of the
X-ray-selected sample of young stars in the CNC due to this foreground cluster
(Sect.~4).

\section{Gaia distances to the massive stars in the CNC\label{sec:OB}}
\subsection{Construction of the OB star sample}

We started composing our sample of massive stars in the Carina Nebula
by using the compilation of 114 O- and B-type (and Wolf-Rayet (WR)) 
stars in the Carina Nebula 
(sorted into the individual stellar clusters Tr~14 -- 16 and Bochum~10 and 11) by 
 \citet{Smith06}.
This sample was complemented by 15 additional OB stars in the Carina Nebula
listed in \citet{2021ApJ...914...18S}.
We also added 8 OB stars which were recently identified in the cluster Tr16-SE 
\citep{Tr16-SE-KMOS}, 
the 5  OB stars known in NGC~3324 \citep[see][]{Preibisch14}, 
and 99 B stars from NGC~3293 \citep{Evans2005}. 
This results in a sample of 241  OB stars in the CNC.

\subsection{Determination of Gaia EDR3 parallaxes and bias correction}

We found Gaia EDR3 matches for all OB stars within a one arcsecond radius;
all but one star (CPD $-59~2636)$ have parallax measurements. 
\citet{2021A&A...649A...4L} found that the parallaxes in Gaia 
EDR3 exhibit an offset to their true value and are therefore biased.
These authors also published an algorithm that can be used to calculate the 
parallax offset for each star as a function of several parameters. 
We made use of this algorithm to determine the  
correction\footnote{The correction is only possible for stars with Gaia EDR3 magnitudes
 between $G=6$~mag  and $G=21$~mag. 
For stars with 5-parameter solutions, the effective wavenumber, i.e.,~the ``photon flux-weighted inverse wavelength as estimated from the BP and RP bands'' \citep{Chapter13-2021}, has to be between 1.1 and 1.9~$\mu \mathrm{m}^{-1}$.
For stars with 6-parameter solutions, the pseudocolor, i.e.,~the ``astrometrically estimated effective wavenumber of the photon flux distribution in the astrometric (G) band'' \citep{Chapter13-2021}, 
has to be between 1.24 and 1.72~$\mu \mathrm{m}^{-1}$ \citep{2021A&A...649A...4L}.
Four of our target stars have an effective wavenumber outside of this interval 
and one (HD~92964) has a G-magnitude smaller than 6~mag.}
for the parallaxes for all but five stars in our sample.

 \begin{figure*}
\centering  
   \includegraphics[width=15.5cm]{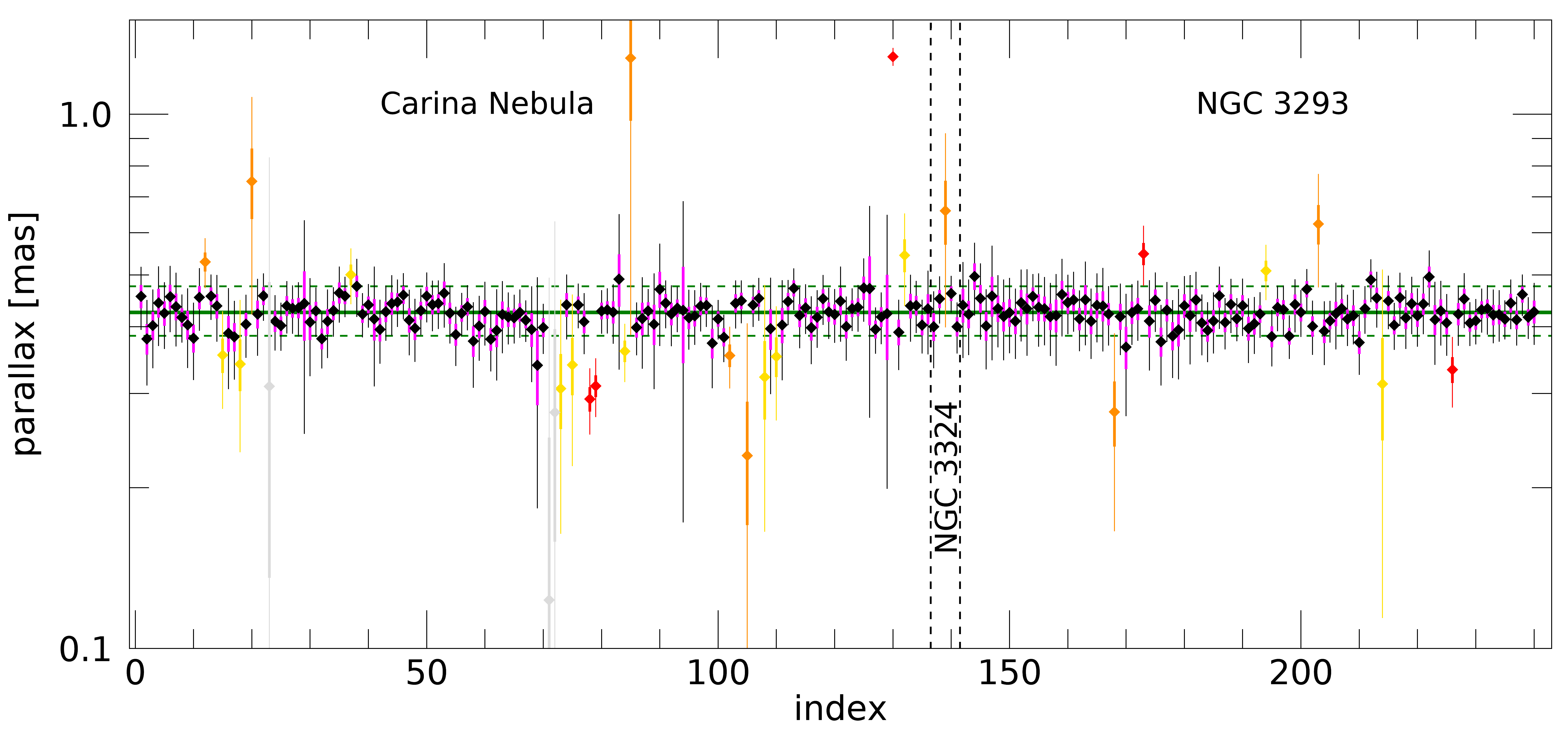}
    \caption{Parallaxes of the OB stars in the different parts of the CNC, shown 
as black diamonds. 
The errorbars in magenta and black show the $1\sigma$ and $3\sigma$ uncertainty of the parallax, respectively.
The green horizontal line shows the parallax at 2.35~kpc, 
while the green dotted lines delimit the 2.1~kpc to 2.6~kpc range. 
The dashed vertical lines delimit the different sub-samples,
i.e.,~the central Carina Nebula region (index 1--136), NGC~3324 (index 137--141),
and NGC~3293 (index 142--240).
The stars which are classified as foreground or background stars 
according to the ``$3 \sigma$'', the ``$2 \sigma$'', and the ``$1 \sigma$'' criterion 
described in Sect.~\ref{sec:outliers} are marked in red, orange, and yellow, respectively.
Stars with a relative parallax uncertainty  $\sigma_\varpi / \varpi \ge 0.3$ 
are shown in light gray. The index numbers on the x-axis refer to the 
number of the stars in the electronic Table 1 (available at the CDS).}
    \label{fig:parOB}
\end{figure*}

The offset values for the stars in our OB sample range from $-0.061686$~mas  
to $-0.010243$~mas. The correction is applied by subtracting these offset values
from the parallaxes listed in the catalog, namely, by~calculating the corrected parallax as
$\varpi_{\mathrm{corr}} = \varpi_{\mathrm{catalog}} - \mathrm{offset}$.
The list of the 240 OB stars with their positions, names,
corrected parallaxes, and spectral types
is given in the electronic Table 1
(available at the CDS).
The corrected parallaxes with their error bars
are plotted in Fig.~\ref{fig:parOB}. 
Almost all stars in the sample have small 
fractional parallax uncertainties of $\sigma_\varpi / \varpi < 0.2$;
only three stars\footnote{HD~93403~A, J10452860-5947553 and J10453468-5947536} have
 high parallax uncertainties of $\sigma_\varpi / \varpi > 0.3$
and they were thus excluded from further analysis.
Figure~\ref{fig:parOB} shows that the large majority of the stars
have very similar and well consistent parallaxes corresponding to distances in the range
2.1~kpc to 2.6~kpc.
A few stars are more or less obvious outliers, that is,~likely foreground or
background stars unrelated to the CNC -- we consider these outliers in Sect.~\ref{sec:outliers}.

For an initial estimate of the distances of the groups, we computed 
the median and the "normalized median absolute deviation"\footnote{The 
\textit{normalized median absolute deviation}
is a robust measure of the variability in a data set
that is scaled to be approximately equal to the standard deviation for
normally distributed data \citep[see, e.g.,][]{FB12}.
}
of the
parallaxes and inverted them to derive distance estimates.
For the OB stars in the Carina Nebula sample, the 
median value of the parallaxes is 0.4242~mas 
and the 
normalized median absolute deviation is $\pm 0.0311$~mas,
corresponding to a median distance of $2.36^{+0.18}_{-0.16}$~kpc.
This value
is in very good agreement\footnote{This is not surprising, since our OB sample
contains all of the stars from \cite{2021ApJ...914...18S}. However, our sample is
considerably larger, and thus provides a useful confirmation of their result.}
 with the result from \citet{2021ApJ...914...18S},
and also in extremely good agreement
to the distance of $(2.35 \pm 0.05)$~kpc for $\eta$~Car
from \citet{Smith06}.

For the  OB stars in NGC~3324 we find  a median parallax of
 $(0.4512 \pm 0.0510)$~mas and corresponding distance of 
    $2.22^{+0.45}_{-0.32}$~kpc,
and for the OB stars in NGC~3293, the results are 
 $(0.4252 \pm 0.0154)$~mas and a corresponding distance of 
$2.35^{+0.13}_{-0.12}$~kpc.
Combining all three sub-samples of OB stars yields a
distance of $2.36^{+0.17}_{-0.14}$~kpc.

\subsection{Identification and exclusion of fore- and background stars \label{sec:outliers} }

In the next step, we take into account that the OB star samples suffer
from a small degree of foreground or background contamination.
As seen in Fig.~\ref{fig:parOB}, several stars
show considerably smaller or larger parallaxes than expected for objects at 
$(2.35 \pm 0.2)$~kpc
and are thus most likely contaminators in the fore- or background.
This is not surprising, since the samples were collected from various
spectroscopic observations of stars in the CNC area and 
it is often difficult to estimate stellar distances from spectra and photometry alone,
especially if the luminosity class of the stars cannot be well measured.

The Gaia parallaxes provide now a way to 
distinguish fore- or background stars from CNC members,  but for this, it is necessary
to define appropriate distance limits.
An inspection of the distribution of parallaxes shows that the large majority
($\simeq 80\%$) of the OB stars show a narrow range of parallaxes 
corresponding to  distances between $\simeq 2.1$~kpc and $\simeq 2.6$~kpc.
We therefore define all the stars for which the 
 full extent of the $3\sigma$ uncertainty range of their parallax corresponds to 
distances of less than 2.1~kpc (i.e.,~$(\varpi - 3 \,\sigma_\varpi) > 1/2.1$~mas)
as clear foreground stars and
all stars for which the full extent of the 
 $3\sigma$ uncertainty range of their parallax corresponds to distances
of more than 2.6~kpc (i.e.~$(\varpi + 3 \,\sigma_\varpi) < 1/2.6$~mas)
as clear background stars.
The five clear fore- or background stars\footnote{Among the stars listed in
\citet{Smith06} as members of the Carina Nebula,
HDE~302989 (SpT = B2~V) is a  foreground star ($1/\varpi = 781$~pc),
HD~93342 (SpT = B1~Ia) is a background star ($1/\varpi = 3.41$~kpc), and
Tr15-18 (SpT = B1~Ia) is a background star ($1/\varpi = 3.23$~kpc).}
 identified by this criterion
in our OB star sample
are highlighted in red in Fig.~\ref{fig:parOB}.
 
The choice of the conditions for excluding stars
introduces some degree of subjectivity into the analysis.
In order to investigate how strongly our choice affects the results,
we also created samples where all stars that are at least $2\sigma$
or at least $1\sigma$ above or below the $1/2.6$ to  $1/2.1$~mas
parallax interval are excluded.
This $2\sigma$-criterion excludes 13 stars\footnote{We note that also one of the
O stars in NGC~3324, HD~92\,206~C (= CD $-57$\,3378, spectral type O7.5~V), 
has a discrepant parallax
that qualifies it as a ``$2 \sigma$ outlier'' foreground stars, although
its location in the center of the Gum~31 nebula seems to suggest
that this is one of the ionizing stars of the Gum~31 \ion{H}{II} region.
HD~92206~C is a double-lined, 2.02 day period spectroscopic binary with a B0~V
companion \citep{2007soch.conf...63C} and a
third component \citep{2017A&A...600A..33M}.
A further possible companion star was found at a separation of $1.7''$
in the \textit{Chandra} X-ray image \cite[see][]{Preibisch14}.
Perhaps the orbital motions in this  multiple
system might have affected the parallax measurement.
This hypothesis is supported by the fact that the Gaia
Renormalised Unit Weight Error (RUWE) of this star has a value
of 5.27, considerably larger than most other stars in our sample. 
Such a high RUWE value ``could indicate that the source is 
non-single or otherwise problematic for the astrometric solution''
(\url{https://gea.esac.esa.int/archive/documentation/GDR2/Gaia_archive/chap_datamodel/sec_dm_main_tables/ssec_dm_ruwe.html}).}, 
and the $1\sigma$-criterion excludes
24 stars from our sample.
In the following distance analysis steps, we compare the effects of the
three different exclusion limits in order to see how robust the results are.

\subsection{Maximum-likelihood analysis of the mean distances}
\label{sec:distOB-ML}

We used a maximum likelihood (ML) procedure to estimate the distance to 
different structures in the CNC.
Assuming a spherical symmetric cluster of stars at a mean distance $D_0$ and
a line-of-sight extent much smaller than the mean distance,
and measurement errors that follow a Gaussian distribution,
the likelihood for obtaining the observed set of parallax values
 $\lbrace \, \varpi_i \, \rbrace$ for $i =  \lbrace 1, \dots , N \rbrace$ with 
associated uncertainties $\lbrace \, {\sigma_{\!\varpi}}_i \, \rbrace$
is given by the product of the $N$ Gaussian functions around the parallax value 
$\varpi_0 = 1 / D_0$:
\begin{equation}
  P(\varpi_0 = 1 / D_0) = \prod_{i=1}^{N} \frac{1}{\sqrt{\,2\pi}\;{\sigma_{\!\varpi}}_i}\;\mathrm{exp}\left(-\frac{\left(\varpi_i-\varpi_0\right)^2}{2\,({\sigma_{\!\varpi}}_i)^2}\right)
\label{eq:ml}
.\end{equation}
The most likely value for the cluster parallax, $\varpi_{\rm ML}$, and the corresponding 
mean distance, $D_{\rm ML} = 1 / \varpi_{\rm ML}$, of the cluster can then
be found by maximizing the likelihood function $P$.
In the Gaussian case of Equation 1, the
 maximum likelihood estimator of the parallax, $\varpi_{\rm ML}$, is simply the 
weighted arithmetic mean of all measurement values with weights of
$w_i = [{\sigma_{\!\varpi}}_i]^{-2}$ 
and the uncertainty of this ML parallax value,
${\sigma_{\!\varpi}}_{\rm ML}$, can also be simply calculated from the sum of the weights.
We note that this uncertainty corresponds to the uncertainty of the mean value,
not the sample standard deviation (as reported, e.g.,~in the results of
\citet{2019ApJ...870...32K}, \citet{Povich19}, and \citet{2021ApJ...914...18S}).

Finally, we compute an estimate of the most likely mean distance, $D_{\rm ML}$, 
and its uncertainty
range by inverting the ML parallax values. 
In general, the inversion of the mean parallax may lead to a biased estimate
of the true mean distance. Since measured 
parallaxes are never error-free, 
and because the transformation from parallax, $\varpi,$ to distance, $D,$ is non-linear, 
the expectation value of $1 / \varpi$ is generally not an unbiased estimate of the 
true distance \citep[see, e.g.,][]{2012aamm.book..242B,2015PASP..127..994B,2018A&A...616A...9L}.
However, for moderate fractional parallax uncertainties ($\sigma_\varpi / \varpi \la 0.3$)
and for a spherical homogeneous group of stars,
the inverse of the mean parallax is still a good estimator of the mean distance of the group
\citep[as long as the uncertainties of the parallax values
are normally distributed; see][]{1999AJ....117..354D}.
Since for almost all stars in our sample, the
fractional parallax uncertainties are small ($\sigma_\varpi / \varpi < 0.2$),
the inversion of the mean parallax should yield a useful distance estimate.
Furthermore, we also estimate distances via a more sophisticated 
Bayesian inference model, which  we describe in Sect.~\ref{sec.Kalkayotl}.

The maximum likelihood mean parallax and distance values for all three
choices of the exclusion criterion are summarized in Table~\ref{table:distances}.
The numbers show that
the different values of the exclusion criterion (i.e.,~$3\sigma$, $2\sigma$, or $1\sigma$)
affect the
resulting mean parallaxes  only very slightly; the corresponding distances
vary by only a few parsec, namely,~$\leq 0.3\%$ of the mean distance.
This directly confirms that our analysis results are robust and do not depend
noticeably on the assumptions established for the exclusion of outlier stars.

Comparing the distances of the three OB star samples,
we find that the ML distances for the OB samples in the Carina Nebula and NGC~3293
are almost identical. The ML distance of NGC~3324 appears slightly ($\sim 60$~pc, i.e.~2.5\%)
 smaller than those for the Carina Nebula and NGC~3293; but due to the
small sample size of only five stars, the uncertainties in the NGC~3324
distance estimate are relatively large,  and it is still consistent with
a common distance for all three samples.
Combining all three OB samples yields a value of $ 2.35^{+0.02}_{-0.02}$~kpc as our best estimate of the maximum likelihood mean distance of the CNC OB stars.

We also investigated how the distance results would change if
we applied a quality criterion on the parallaxes, by excluding stars
for which the ``renormalised unit weight error'' \citep[RUWE, an indicator
for the quality of the astrometric solution; see][]{LL:LL-124} is larger than 1.4.
All details about these calculations and the resulting distance values
 are described in Appendix~1. We find that
applying this  quality criterion leads only to very marginal changes
of the distances by just a few pc and this would neither affect our conclusions,
nor the final distance estimates quoted here (by three significant digits)
in any way.

\subsection{Bayesian inference of distances with \textit{Kalkayotl} \label{sec.Kalkayotl}}

As a second way to determine distances we employed the program \textit{Kalkayotl}
developed by \citet{2020A&A...644A...7O}.
\textit{Kalkayotl} is a free and open code that
uses a Bayesian hierarchical model to obtain samples of the
posterior distribution of the cluster mean distance by means of a Markov
chain Monte Carlo (MCMC) technique implemented in PyMC3.
\textit{Kalkayotl} also takes the parallax spatial correlations into account,
which improves the credibility of the results.
\citet{2020A&A...644A...7O} performed extensive tests and found that 
\textit{Kalkayotl} allows for trustworthy estimates of 
cluster distances to be derived up to about 5~kpc from Gaia data.

We used \textit{Kalkayotl} version 1.1,
with the parameters generally left at their standard values.
For the prior, we used the implemented  Gaussian model with a mean distance of
$D_{\rm prior} = (2.35 \pm 0.1)$~kpc and a cluster scale 
of $S_{\rm prior} = 100$~pc. The calculations were done in distance space, and
the reported uncertainties for the inferred mean distances
are the central 68.3\% quantiles\footnote{corresponding to the ``$\pm 1 \sigma$
range'' for a Gaussian distribution}.
The resulting distances are also listed in Table~\ref{table:distances}
and in Table~A.1 for the distances determined with the RUWE selection
criterion.

\subsection{Discussion of distance results}

\begin{table*}  \setcounter{table}{1}
\caption{Maximum likelihood and \textit{Kalkayotl} estimates for the distances}
\label{table:distances}
     \centering
\begin{tabular}{l|rrr|crc|rc}
\hline\hline                 
  Group &  \multicolumn{3}{c|}{outlier exclusion:} & ML parallax &
\multicolumn{2}{c}{ML distance}  &
\multicolumn{2}{|c}{\textit{Kalkayotl} distance}   \\
&  & {\small fore-} & {\small back-} & $\varpi_{\rm ML}$ & $D_{\rm ML}$
&  $2 \sigma$-range & $D_{\rm Kal}$ &  {\small central 68.3\% quant.}\\
   &  & \multicolumn{2}{c|}{{\small ground}} & [mas] &
\multicolumn{2}{c}{[kpc]} & \multicolumn{2}{|c}{[kpc]} \\
\hline                        
Carina Neb.  & $3\sigma$ & 1 & 2 & $0.42440\pm 0.00153$ & $2.356$ &$[ 2.339 \, , \, 2.373 ]$ & 2.367 & $[ 2.312 \, , \, 2.422 ]$\\
OB stars     & $2\sigma$ & 4 & 4 & $0.42444\pm 0.00155$ & $2.356$ &$[ 2.339 \, , \, 2.373 ]$ & 2.367 & $[ 2.312 \, , \, 2.421 ]$ \\
 $(N = 133)$ & $1\sigma$ & 6 & 11& $0.42548\pm 0.00157$ & $2.350$ &$[ 2.333 \, , \, 2.368 ]$ & 2.363 & $[ 2.307 \, , \, 2.418 ]$\\
\hline                        
  NGC~3324 & $3\sigma$ &  0 & 0 & $0.43745\pm 0.00733$ & $2.286$ &$[ 2.212 \, , \, 2.365 ]$ &2.321 & $[ 2.250 \, , \, 2.393 ]$ \\
  OB stars & $2\sigma$ &  1 & 0 & $0.43585\pm 0.00735$ & $2.294$ &$[ 2.219 \, , \, 2.374 ]$ &2.338 & $[ 2.268 \, , \, 2.410 ]$ \\
 $(N = 5)$ & $1\sigma$ &  1 & 0 & $0.43585\pm 0.00735$ & $2.294$ &$[ 2.219 \, , \, 2.374 ]$ &2.338 & $[ 2.268 \, , \, 2.410 ]$ \\
\hline                        
  NGC~3293 & $3\sigma$ & 1 & 1 & $0.42523\pm 0.00178$ & $2.352$ & $[ 2.332 \, , \, 2.372 ]$ & 2.354 & $[ 2.299 \, , \, 2.410 ]$\\
OB stars   & $2\sigma$ & 2 & 2 & $0.42531\pm 0.00178$ & $2.351$ & $[ 2.332 \, , \, 2.371 ]$ & 2.356 & $[ 2.299 \, , \, 2.412 ]$ \\
$(N = 99)$ & $1\sigma$ & 3 & 3 & $0.42471\pm 0.00179$ & $2.355$ & $[ 2.335 \, , \, 2.375 ]$ & 2.357 & $[ 2.301 \, , \, 2.414 ]$\\
\hline                        
All        & $3\sigma$ & 2 & 3 & $0.42506\pm 0.00115$ & $2.353$  & $[ 2.340 \, , \, 2.365 ]$ & 2.363 & $[ 2.308 \, , \, 2.417 ]$\\
  OB stars & $2\sigma$ & 7 & 6 & $0.42509\pm 0.00115$ & $2.352$  & $[ 2.340 \, , \, 2.365 ]$ & 2.363 & $[ 2.309 \, , \, 2.418 ]$ \\
$(N = 237)$& $1\sigma$ &10 &14 & $0.42542\pm 0.00117$ & $2.351$  & $[ 2.338 \, , \, 2.364 ]$ & 2.362 & $[ 2.308 \, , \, 2.416 ]$\\
\hline\hline
CCCP         & $3\sigma$ &  906 &  217 & $0.43097\pm 0.00065$ & $2.320$ & $[ 2.313 \, , \, 2.327 ]$ & 2.337 & $[ 2.284 \, , \, 2.391 ]$\\
X-ray        & $2\sigma$ & 1171 &  577 & $0.43015\pm 0.00067$ & $2.325$ & $[ 2.318 \, , \, 2.332 ]$ & 2.336 & $[ 2.281 \, , \, 2.389 ]$\\
$(N = 8505)$ & $1\sigma$ & 1908 & 1523 & $0.43005\pm 0.00069$ & $2.325$ & $[ 2.318 \, , \, 2.333 ]$ & 2.330 & $[ 2.276 \, , \, 2.384 ]$\\
\hline                        
NGC~3324    & $3\sigma$ &  61 & 13 & $0.42329\pm 0.00278$ & $2.362$ & $[ 2.332 \, , \, 2.394 ]$ & 2.348 & $[ 2.289 \, , \, 2.407 ]$\\
X-ray       & $2\sigma$ &  76 & 29 & $0.42145\pm 0.00282$ & $2.373$ & $[ 2.341 \, , \, 2.405 ]$ & 2.358 & $[ 2.301 \, , \, 2.416 ]$\\
$(N = 455)$ & $1\sigma$ & 113 & 64 & $0.42282\pm 0.00300$ & $2.365$ & $[ 2.332 \, , \, 2.399 ]$ & 2.361 & $[ 2.304 \, , \, 2.418 ]$\\
\hline                        
NGC~3293    & $3\sigma$ &  58 & 10 & $0.43130\pm 0.00216$ & $2.319$ & $[ 2.296 \, , \, 2.342 ]$ & 2.346 & $[ 2.290 \, , \, 2.402 ]$\\
X-ray       & $2\sigma$ &  73 & 32 & $0.43165\pm 0.00217$ & $2.317$ & $[ 2.294 \, , \, 2.340 ]$ & 2.342 & $[ 2.286 \, , \, 2.398 ]$ \\
$(N = 602)$ & $1\sigma$ & 121 & 90 & $0.43073\pm 0.00222$ & $2.322$ & $[ 2.298 \, , \, 2.346 ]$ & 2.342 & $[ 2.287 \, , \, 2.398 ]$ \\
\hline                        
All         & $3\sigma$ & 1025 &  240 & $0.43063\pm 0.00061$ & $2.322$ & $[ 2.316 \, , \, 2.329 ]$ & 2.338 & $[ 2.286 \, , \, 2.391 ]$\\
X-ray       & $2\sigma$ & 1320 &  638 & $0.42985\pm 0.00062$ & $2.326$ & $[ 2.320 \, , \, 2.333 ]$ & 2.336 & $[ 2.282 \, , \, 2.390 ]$ \\
$(N = 9562)$& $1\sigma$ & 2142 & 1677 & $0.42977\pm 0.00065$ & $2.327$ & $[ 2.320 \, , \, 2.334 ]$ & 2.331 & $[ 2.279 \, , \, 2.384 ]$ \\
\hline                                   
\end{tabular}
\end{table*}

The comparison of the distances derived with the ML procedure and
with  \textit{Kalkayotl} shows generally good agreement; in most cases,
the differences are no more than 10--20~pc, 
that is, less than 1\% of the absolute distance, showing that our distance results 
are robust with respect to the employed analysis method.
Comparing the three OB samples, also the \textit{Kalkayotl} results indicate
that the five NGC~3324 OB stars are slightly closer than the mean distance for the 
Carina Nebula and NGC~3293 OB stars, but this small difference is not significant
and may well be purely due to the small sample size.

Combining all three OB star samples, we find a mean distance
 of $2.36^{+0.05}_{-0.05}$~kpc for the OB stars in the CNC.
This result is in excellent agreement
to the previously determined distance 
of $(2.35 \pm 0.05)$~kpc found by \citet{Smith06} for $\eta$~Car.

\section{Gaia distances to the X-ray detected stars in the CNC}

Although the OB stars considered in the previous section are the most
prominent objects, the vast majority of stars in the complex are 
low-mass stars. But the population of low-mass stars in the CNC is still
only poorly known, since spectroscopic observations of these faint stars
are very time-consuming and have been obtained thus far for a very small 
fraction of the low-mass stars
in the CNC. Due to the celestial position of the Carina Nebula 
very close to the Galactic plane,
there is also a very high rate of background contamination 
in the CNC, making it very difficult to identify low-mass (= optically faint) 
CNC members among the millions of Galactic fore- and background stars
in the CNC area \citep[see][]{VISTA1}.

X-ray observations provide a very good way to solve this problem by
picking out young, coronally active stars among the much older
(and usually X-ray faint) Galactic field stars \citep[see, e.g.][]{PF05}.
In the CNC, the combination of the
comprehensive CCCP X-ray survey \citep[see][]{CCCP-intro}
with the individual \textit{Chandra} observations of NGC~3324 \citep{Preibisch14}
and NGC~3293 \citep{2017A&A...605A..85P}
have provided an approximately mass-limited $(M \ga 0.5\, M_\odot$)
sample of the young low-mass stars in the complex.

\subsection{Construction of the X-ray sample}
\label{sec:x-rayconstr}

The CCCP X-ray catalog of 14~368 X-ray sources described in \citet{CCCP-catalog} 
was used as our sample in the Carina Nebula. 
Most of these X-ray sources are 
young stars in the Carina Nebula, but a significant contamination with 
X-ray active foreground-stars and 
background-stars, as well as with extragalactic objects is to be expected \citep[see][and our Sect.~3.4]{CCCP-classification}.
In order to analyze the Gaia EDR3 parallaxes of the X-ray sources,
we matched the CCCP source catalog with Gaia EDR3 sources via the
program match\_xy \citep[see][]{Broos2010}. 
This program is preferred over simple matching algorithms with a fixed
``search radius'' since the positional uncertainties of the X-ray sources 
show a rather wide variation\footnote{X-ray source positional uncertainties
are a function of the source position on the detector and the number of detected
X-ray counts for each source.} and range from
$0.054^{\prime\prime}$ up to $1.918^{\prime\prime}$ in the CCCP sample.
For each pair of possible matches, match\_xy calculates
the probability for the
null hypothesis that the two catalog entries are observations of the same object
by assuming Gaussian positional uncertainties and a significance threshold of 0.99.
The algorithm also enforces a one-to-one relationship among the 
pairs of sources declared to be matches. More detailed information 
about the algorithm is given in \citet{Broos2010}.

The positional uncertainties for the X-ray sources were taken directly from 
the CCCP catalog, while the Gaia EDR3 positional uncertainties were assumed to 
be $0.01^{\prime\prime}$.
The matching procedure revealed a very small systematic positional shift of 
$0.01^{\prime\prime}$ in right ascension and of $0.005^{\prime\prime}$ in declination between the CCCP and Gaia EDR3 positions. 
After correcting this systematic shift, we found 9111 Gaia EDR3 matches to X-ray source positions. 
The X-ray source catalogs of the star clusters 
NGC~3324 \citep[][679 X-ray sources]{Preibisch14} and 
NGC~3293 \citep[][1026 X-ray sources]{2017A&A...605A..85P} were 
matched to Gaia EDR3 in the same way.
This revealed small positional shifts of $0.09^{\prime\prime}$ in RA and $0.015^{\prime\prime}$ in Dec for NGC~3324 and 
$0.11^{\prime\prime}$ in RA and $0.18^{\prime\prime}$ in Dec for NGC~3293.
The number of resulting Gaia EDR3 matches was 480 for NGC~3324 and 648 for NGC~3293.

\subsection{Determination of Gaia EDR3 parallaxes and bias correction}

In the CCCP-region, 8505 X-ray sources with Gaia EDR3 
match have a parallax. 
In NGC~3324 and NGC~3293, the corresponding numbers are 455 
and 602, respectively.
The algorithm provided by \citet{2021A&A...649A...4L} was used again to determine the parallax bias for each source. 
The majority of stars could be corrected via this method (6482 for CCCP, 
319 for NGC~3324, and 498 for NGC~3293).

\subsection{Analysis of the distances of the X-ray-selected stars}

We used the same methods and criteria  for the mean distance determination as in 
our analysis for the OB stars described in Sect.~2. 
Among the 8505 Gaia EDR3 matches with parallax in the CCCP sample,
906 objects are classified as foreground stars and 217 objects as 
background stars based on their parallaxes and the $3 \sigma$-criterion described
in Sect.~\ref{sec:outliers}.

The mean distances derived with the ML method and with \textit{Kalkayotl}
for the three different X-ray samples and the different outlier rejection criteria
are summarized in Tab.~\ref{table:distances}.
The \textit{Kalkayotl} distances are again very similar
to the ML distances (with differences of $\la 20$~pc, i.e.,~$\la 1\%$ of the absolute
distance) for all three samples.
Comparing the three different samples shows again that NGC~3293 and
the Carina Nebula population have almost identical distances. The X-ray-selected
stars in NGC~3324 appear to be marginally ($\approx 40$~pc according to the ML results
and $\approx 20$~pc according to the \textit{Kalkayotl} results) more distant
than the stars in the Carina Nebula, but this difference is not significant and
corresponds to $\la 1.5\%$ of the absolute distance.

\begin{table*}
\caption{Number of X-ray-selected objects with Gaia parallaxes
consistent or not consistent with their photometric classification in
\citet[][\mbox{[B11]}]{CCCP-classification}.}
\label{tab:classification}
\centering
\begin{tabular}{l r r r r}
\hline\hline
Class & number of & Gaia matches & Consistent  with & Not consistent with \\
 & X-ray sources & with parallax & B11 classification  & B11 classification \\
\hline
 1 (foreground) & 716     &  447  & 429 \, (96.0\%) & 18 \, \;\;(4.0\%)\\
 2 (Carina Nebula) & 10\,714 & 7262  & 6571 \, (90.5\%) & 691 \, \;\;(9.5\%)\\
 3 (background) &  16     &   10  & 3 \, (30.0\%) & 7 \, (70.0\%)\\
 4 (extragalactic) & 877     &   89 & 87 \, (97.8\%) & 2 \, \;\;(2.3\%)\\
\hline
\end{tabular}
\end{table*}

The distances derived for the X-ray samples are also generally 
in good agreement with those
for the corresponding OB star samples; in most cases, the differences
are $\la 30 $~pc, namely,~$\la 1.3\%$ of the absolute distance.
We note that while the distance estimates for the OB stars in NGC~3324
were $\sim 30 - 50$~pc lower than those for the Carina Nebula OB stars,
the distance estimates for the  NGC~3324 X-ray sources are
$\sim 20 - 40$~pc higher than those for the Carina Nebula X-ray sources.
This suggests that these small ($\leq 2\%$) differences in the distances
for NGC~3324 and the Carina Nebula are not significant.

These results for the X-ray-selected samples provide strong support for 
the conclusions drawn above from the distance results for the OB star samples.
They confirm that the clusters NGC~3324 and NGC~3293 are 
(within less than 2\%) 
at the same distance as the central Carina Nebula.
Combining all three X-ray samples yields a final value of 
\mbox{$2.34^{+0.05}_{-0.06}$~kpc} as our
best estimate of the mean CNC X-ray star distance, which is very close (only 1\% smaller)
to the  \mbox{$2.36^{+0.05}_{-0.05}$~kpc} for the OB stars in 
the CNC derived above.

\subsection{Fore- and background contamination in the CCCP sample}

Long before the availability of Gaia parallaxes,
\citet[][B11 hereafter]{CCCP-classification} addressed the problem of
Galactic fore- and background contamination of the X-ray-selected sample in the CCCP and
performed a probabilistic classification of the CCCP
X-ray sources that was solely based on the X-ray properties
and the optical and infrared photometry of their counterparts.
They classified the X-ray sources into four classes: 
5.0\% of the X-ray sources were classified as ``likely foreground stars'' (class 1),
74.6\% as ''likely Carina pre-main-sequence stars'' (class 2),
0.1\% as ``likely background stars'' (class 3), and
6.1\%  as ``likely extragalactic sources'' (class 4).
For about 14.2\% of the X-ray objects, the
classification procedure gave no convincing clues and these objects thus 
remained unclassified. 

With the Gaia EDR3 parallaxes available today, we can now compare this photometric probabilistic classification with the new direct distance information
 from the Gaia parallaxes, and check how well the purely photometric probabilistic classification worked.
For this, we assumed a source classification as inconsistent with the Gaia parallax 
 if the $3\sigma$ uncertainty interval lies completely outside 
the distance range corresponding to its
classification, namely,~$<2.1$~kpc for foreground stars, 2.1-- 2.6~kpc
for Carina Nebula stars, and $> 2.6$~kpc for background stars and extragalactic objects.
The results of this comparison are compiled in Table~\ref{tab:classification}. 

The comparison shows that in the large majority (90.8\%) of cases the B11
classification as foreground or background or Carina Nebula member is consistent
with the Gaia parallax of the objects; it is only in 9.2\% of cases that the 
Gaia parallax is inconsistent with the B11 classification.
In addition, it is only in the (very small) group of likely background stars the classification
did not work well, but this is, firstly, not unexpected (since obtaining a purely photometric 
distinction between background objects and strongly obscured stars in the Carina Nebula
is very difficult),
and, secondly, this leads to only seven misclassified objects (i.e.,~a vanishingly small 
fraction of the 14\,368 X-ray sources).
In conclusion, this comparison shows that the B11 probabilistic classification
worked remarkably well in sorting the X-ray sources into  different classes.

Considering the 2045 X-ray sources that remained unclassified in B11,
we found 697 Gaia EDR3 matches with parallaxes. 
The large majority (73.9\%) of these have parallaxes that are consistent 
(within their $3\sigma$ uncertainty interval) with being members of the Carina Nebula. 
The rest are either foreground (23.0\%) or background objects (3.1\%)
according to their parallaxes.

\section{The foreground cluster vdBH~99}

Due to the Carina Nebula's position very close to the Galactic plane,
contamination by field stars and also possibly by stellar clusters in the
fore- or background is a serious issue.
The stellar cluster vdBH~99,
first mentioned by \citet{1975AJ.....80...11V}, is located just about one degree
northwest of the Carina Nebula, and just south of NGC~3324/Gum~31, 
and therefore clearly causes foreground contamination.
The optically brightest cluster member HD~92397 ($m_V = 4.87$~mag) 
is a K4.5 AGB or supergiant star \citep{2019AJ....158...20M}.
Using Gaia DR2 data,
\citet{2018A&A...618A..93C} determined the center of vdBH~99 at the J2000 
coordinates $(\alpha,\delta)=(159.553\degr, -59.168\degr)$ 
and a radius which encloses half of the members of $r_{50}=0.537\degr$.
They also derived a mean distance of $(443.6 \pm 0.3)~$pc for vdBH~99.
\citet{2019A&A...623A.108B} found an age of $81$~Myr for vdBH~99 based
on Gaia DR2 parallaxes and photometry.

Since the fractional X-ray luminosities ($L_{\rm X} / L_{\rm bol}$) of young stars
are approximately constant during the first $100$~Myr \citep[see][]{PF05},
the low-mass stars in vdBH~99 are expected to be relatively bright X-ray sources
and to contaminate the X-ray-selected samples of young stars in the CNC to some
degree.
With the Gaia data, we can now identify members of vdBH~99 by their
parallaxes and proper motions, and quantify the
level of foreground contamination for the CNC.

\subsection{Selection of cluster members with Gaia EDR3 data}

\begin{figure} 
\includegraphics[width=8.5cm]{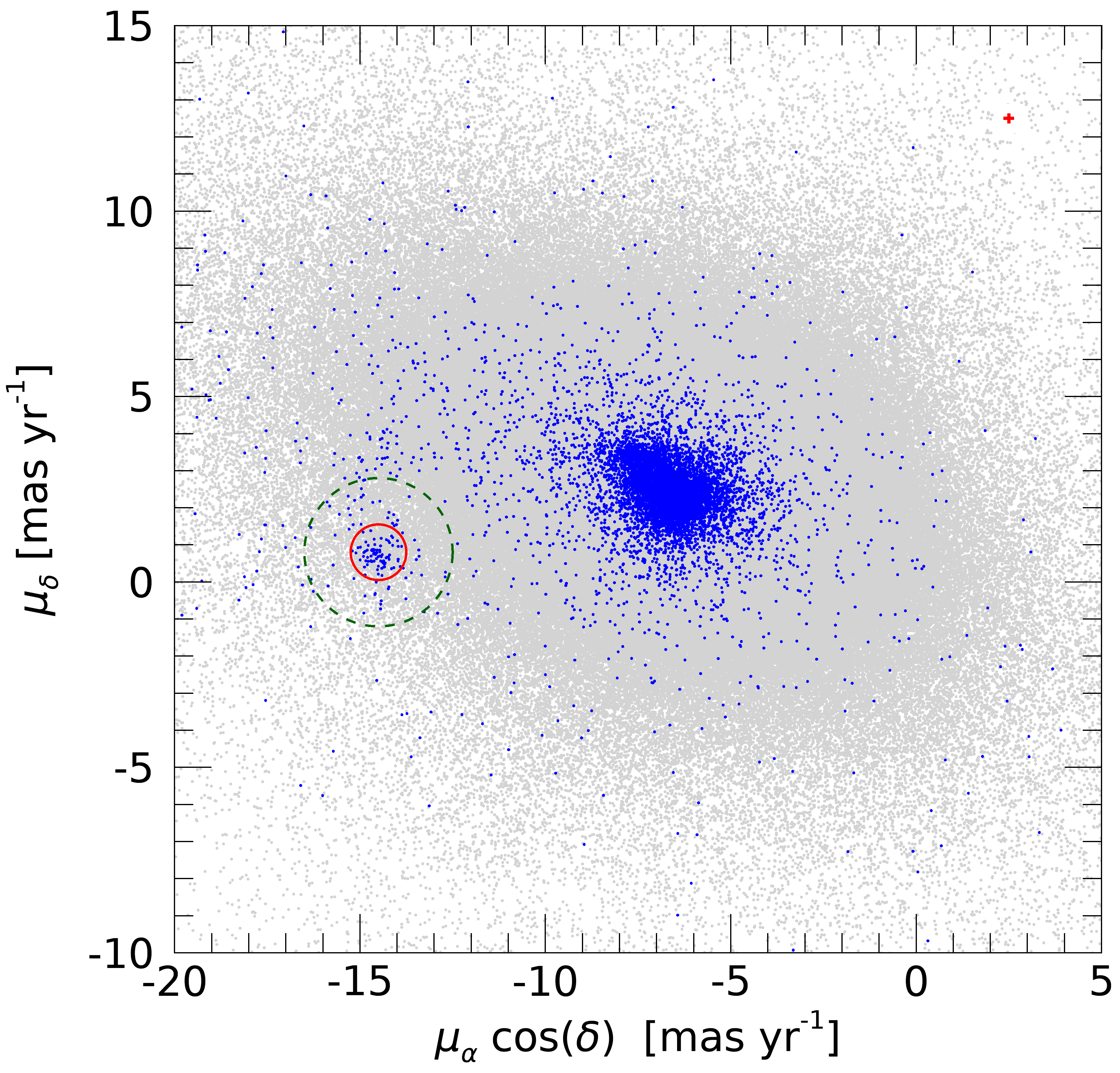}
    \caption{Proper-motion diagram for the CNC region. 
The grey dots show all Gaia EDR3 objects in a $4.75\degr \times 3\degr$ box centered 
around the J2000 celestial coordinates  $ {\rm RA} =10^{\mathrm{h}}\,43^{\mathrm{m}}$, ${\rm Dec} = -59\degr\,20^{\prime}$. 
The blue dots show the proper motions of the X-ray-selected objects of the CCCP, 
NGC~3324, and NGC~3293. The green and red circles mark the area in proper motion
space used for building the histograms of the parallaxes shown in Fig.~\ref{fig:parallax-histogram}.
The red cross in the upper right edge of the plot shows
median values [(0.101, 0.092)~mas\,yr$^{-1}$] of the  
proper motion uncertainties for the stars selected as vdBH~99 members.}
    \label{fig:pm}
\end{figure}

\begin{figure}
\includegraphics[width=8.5cm]{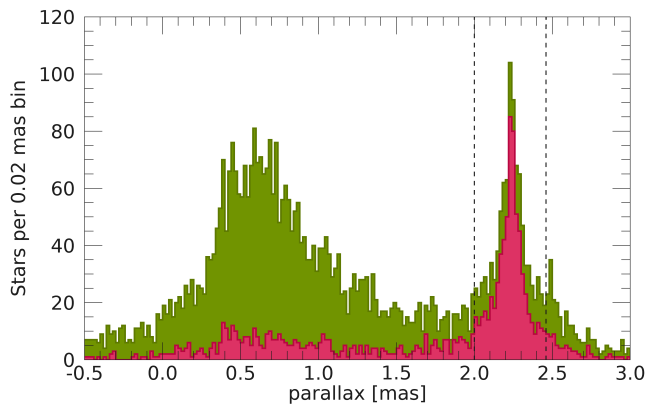}
    \caption{Histograms of the parallaxes of
all Gaia EDR3 objects in a $4.75\degr \times 3\degr$ box centered
around the J2000 celestial coordinates  
$ {\rm RA} =10^{\mathrm{h}}\,43^{\mathrm{m}}$, ${\rm Dec} = -59\degr\,20^{\prime}$
and with proper-motion vectors within
$2\;\mathrm{mas}\,{\mathrm{yr}^{-1}}$ (green) or
$0.75\;\mathrm{mas}\,{\mathrm{yr}^{-1}}$  (red) around $(\mu_{\alpha^*}, 
\mu_{\delta})=(-14.5, 0.8)\;\mathrm{mas}\,{\mathrm{yr}^{-1}}$.
The dashed vertical lines mark the boundaries of the parallax selection interval
for vdBH~99 members.}
    \label{fig:parallax-histogram}
\end{figure}

In Fig.~\ref{fig:pm}, we show a proper motion diagram of all Gaia EDR3 objects in 
and around the CNC,
where the X-ray detected objects are marked by color.
It can be seen that the large majority of X-ray-selected objects in the CNC
show proper motions that are basically indistinguishable from the
proper motions of the galactic field population at this celestial position,
centered roughly around $(\mu_{\alpha^*}, \mu_{\delta}) \approx (-6.5, 2.5)\;\mathrm{mas}\,{\mathrm{yr}^{-1}}$.
However, Fig.~\ref{fig:pm} also shows a clear group of X-ray detected stars 
with proper-motions around
$(\mu_{\alpha^*}, \mu_{\delta})=(-14.5, 0.8)\;\mathrm{mas}\,{\mathrm{yr}^{-1}}$ that is quite  distinct from the CNC and galactic field population.

Figure~\ref{fig:parallax-histogram} shows the distribution of parallaxes of all 
Gaia stars in and around the CNC
with proper-motion vectors within $2~\,\mathrm{mas}\,\mathrm{yr}^{-1}$ 
or $0.75\;\mathrm{mas}\,{\mathrm{yr}^{-1}}$ around 
$(\mu_{\alpha^*}, \mu_{\delta})=(-14.5, 0.8)\;\mathrm{mas}\,\mathrm{yr}^{-1}$
(i.e., the regions marked by the green and red circles in Fig.~\ref{fig:pm}).
Besides the broad peak of parallaxes between $\sim 0.3$~mas and $\sim 1$~mas,
which is produced by the galactic field population and the CNC members,
the distribution of parallaxes also shows another clear peak 
at $\varpi \simeq 2.23$~mas,
which contains the likely members of vdBH~99.
We define the ranges of the parallax selection interval for vdBH~99 members
as $\pm 0.23$~mas  ($\simeq 10\%$ of the mode value),
namely,~extending from 2.0~mas to 2.46~mas.

Since a large fraction of the stars in the $2~\,\mathrm{mas}\,\mathrm{yr}^{-1}$
proper-motion selection circle are not vdBH~99 members but unrelated 
field stars, we tried to find an efficient selection criterion 
for vdBH~99 members in proper motion space. For this,
we considered different radii 
for the proper motion selection circle and 
tested how many, and what fraction of the X-ray-detected stars in these circles,
have parallaxes between 2.0 and 2.46~mas; that is, which ones~might be considered 
 members of the foreground cluster.
We tried proper motion selection circle radii ranging from  
$0.25\,\mathrm{mas}\,\mathrm{yr}^{-1}$ to $2\,\mathrm{mas}\,\mathrm{yr}^{-1}$,
and we found that a radius of $0.75\;\mathrm{mas}\,\mathrm{yr}^{-1}$
yielded the highest fraction (92\%) of stars in the
2.0--2.46~mas parallax interval. This choice provides a good compromise between a 
high level of completeness for likely vdBH~99 members
and a low level of contamination by unrelated field stars.

For the final selection of likely vdBH~99 cluster members, we
used the following selection criteria: 
(1)
a sky position within a $2\degr$ radius around $(\alpha,\delta)=(159.553, -59.168)\degr$ 
which is the center of vdBH~99 as determined by \citet{2018A&A...618A..93C};
(2) a proper motion vector inside a circle with a radius of $0.75\;\mathrm{mas}\,{\mathrm{yr}^{-1}}$ 
centered around 
$(\mu_{\alpha^*}, \mu_{\delta})=(-14.5, 0.8)\;\mathrm{mas}\,{\mathrm{yr}^{-1}}$, as indicated by the red circle in Fig.~\ref{fig:pm}; 
(3) a parallax between 2.0 and 2.46~mas,
 with a relative uncertainty of $\sigma_\varpi / \varpi < 0.2$.
We found 700 stars in the Gaia EDR3 catalog satisfying all three criteria.
 
In order to estimate the completeness of our selection for
vdBH~99 cluster members, we analyzed the distribution of all
Gaia stars in the $2\degr$ radius circle on the sky and
the selected parallax interval in the $(\mu_{\alpha^*}, \mu_{\delta})$ proper motion space.
We found that the peak of the distribution of proper motions, 
$\mu_{\alpha^*}$, around $\simeq -14.5\;\mathrm{mas}\,{\mathrm{yr}^{-1}}$
and
$\mu_{\delta}$ around $\simeq 0.8\;\mathrm{mas}\,{\mathrm{yr}^{-1}}$
can both be approximately characterized by Gaussian distributions with a
standard deviation of 
$\sigma_{\mu_{\alpha^*}} \simeq \sigma_{\mu_{\delta}} \simeq 0.37\;{\rm mas}\,{\rm yr}^{-1}$.
Since our proper motion selection circle radius of $0.75\,{\rm mas}\,{\rm yr}^{-1}$
is just  $\simeq 2$ times the proper-motion standard deviations  
$\sigma_{\mu_{\alpha^*}} \simeq \sigma_{\mu_{\delta}}$,
we estimate the completeness\footnote{For a two-dimensional Gaussian distribution with
equal standard deviation in x- and y-direction ($\sigma_x = \sigma_y = \sigma_{1\rm D}$),
a circle with a radius of $2\,\sigma_{1\rm D}$  contains 86.5\% of the values.}
 of our proper-motion selection to be $\approx 85\%$.

\begin{figure}
\includegraphics[width=8.85cm,trim=50 230 230 280,clip]{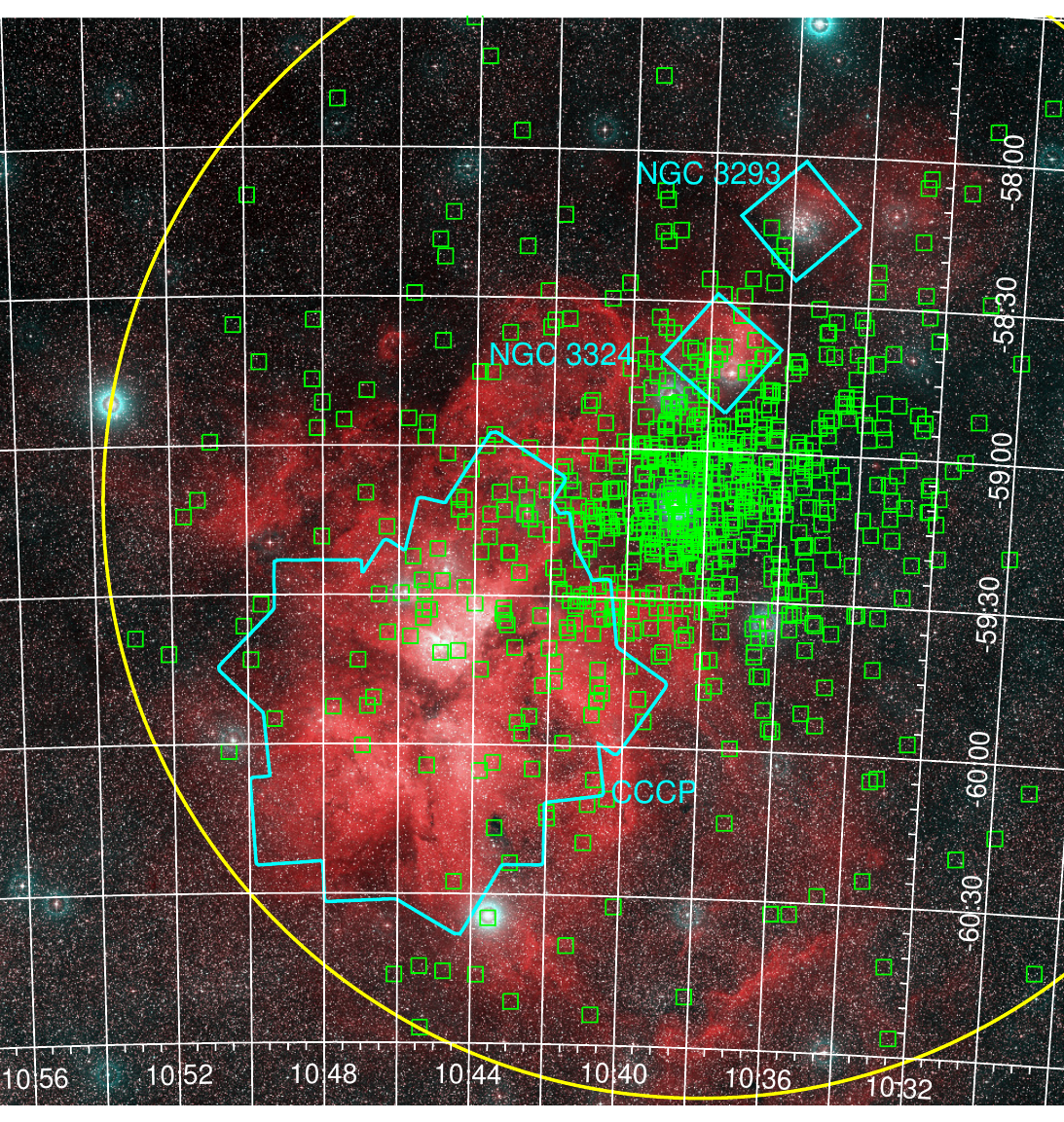}
\caption{Overlap between \mbox{vdBH~99} and the CNC. The background image is an 
optical image from the ESO Photo Release eso0905b
(image credit: ESO/Digitized Sky Survey 2, Davide De Martin).
The proper motion and parallax selected members of vdBH~99 are represented as green boxes, the circle shows the $2\degr$ search radius.}
    \label{fig:vdBH99-DSS}
\end{figure}

\subsection{Determination of cluster properties}

The spatial distribution of the  likely vdBH~99 cluster members 
is shown in Fig.~\ref{fig:vdBH99-DSS}. 
Using the bias-corrected parallaxes 
and the maximum-likelihood procedure described above, we found 
a mean parallax of $(2.2734 \pm 0.0104)$~mas for vdBH~99,
corresponding to a mean distance of $(439.9 \pm 0.2)~$pc.
The distance determined with \textit{Kalkayotl} is $440.8^{+2.3}_{-2.3}$~pc.
Our values are  marginally smaller than the $(443.6\pm0.3)~$pc determined by 
\citet{2018A&A...618A..93C} based on Gaia DR2 data. 
  
\begin{figure}
\includegraphics[width=8.5cm]{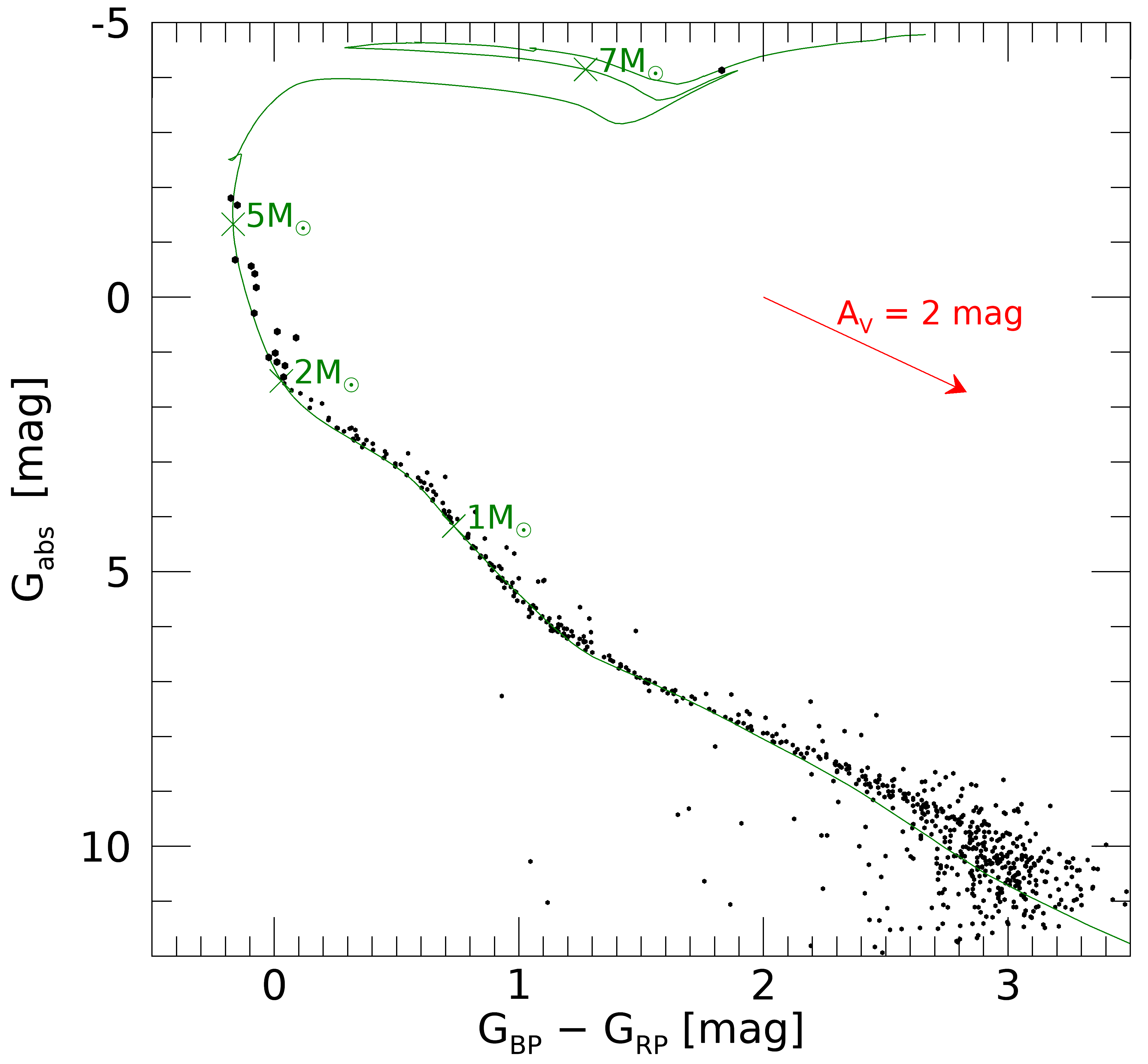}
    \caption{Color-magnitude diagram for members of the cluster vdBH~99. 
The green line shows the PARSEC isochrone for an age of $60$~Myr
 with an extinction of $A_V=0.2$~mag. The red arrow indicates the direction
of the extinction vector.}
    \label{fig:isochrone}
\end{figure}

We also determined the age of vdBH~99 by fitting theoretical isochrones to the 
distribution of  likely members in the Gaia $G$ versus $G_{\rm BP} - G_{\rm RP}$
color-magnitude diagram shown in Fig.~\ref{fig:isochrone}. 
The isochrones were taken from the current version of the 
PARSEC\footnote{available at \url{http://stev.oapd.inaf.it/cgi-bin/cmd_3.4}}
 tracks \citep{2012MNRAS.427..127B}.
We obtained the best fit with an age of $60$~Myr and an extinction of 
$A_V = 0.2$~mag. 
This is somewhat younger than the 81~Myr age estimation by \citet{2019A&A...623A.108B},
which was based on Gaia DR2 data and an assumed extinction of $A_V = 0.203$~mag.

\subsection{Contamination of the X-ray samples in the CNC by vdBH~99 stars}

Finally, we investigated the contamination of the samples of X-ray sources in the CNC
by members of vdBH~99. 
Using the selection criteria described in Sect.~4.1, we
found 89  likely vdBH~99 members that have matches to an X-ray source
in the CNC.
As expected, the contamination rate depends on position: it is very low with
$67 / 14\,368 = 0.47\%$ for the CCCP sample and $ 5 / 1026 = 0.49\%$ for the NGC~3293 
X-ray sample, but $17 / 679 = 2.5\%$ for the NGC~3324 X-ray sample, which is
closest on the sky to the center of vdBH~99.
Even if we consider that our sample of likely vdBH~99 members is not
fully complete, these numbers suggest
that the overall 
level of contamination of the X-ray sources in the CNC by foreground members of vdBH~99
is quite small and does not affect
the conclusions drawn about the young stellar populations in the CNC in our 
earlier studies of the X-ray-selected populations
\citep{CCCP-HAWKI,Preibisch14,2017A&A...605A..85P}.

\section{Conclusions and summary}

Our analysis of the Gaia EDR3 parallaxes for the samples of massive stars
and X-ray detected stars in
the (inner) Carina Nebula, NGC~3324, and NGC~3293 shows
that these star samples are at a common, well-constrained distance
of  $2.35^{+0.05}_{-0.05}$~kpc; 
any differences in the derived cluster distances 
are insignificant and smaller than 2\% of the absolute distance.
The Gaia-based distance we find here
agrees very well with the distance to $\eta$~Car
derived by \citet{Smith06} (by the expansion of the Homunculus Nebula)
and with the old ``canonical'' distance of $(2.3 \pm 0.1)$~kpc
motivated by \citet{SB08}, which had been adopted in many recent studies
of the CNC.

Our result that NGC~3324 and NGC~3293 at the northwestern periphery of the CNC
 are (within $\la 2\%$) at the same
distance as the star clusters in the central Carina Nebula, 
clearly refutes some (pre-Gaia) claims in the literature 
about vastly different distances among these clusters.
The Gaia EDR3 data show  clearly
that the CNC is a coherent physical structure at a common distance,
not a chance projection of unrelated stellar clusters at
different distances.

Since the conclusions drawn in previous investigations about the
ages and the mass functions of the stellar clusters in the CNC
are strongly dependent on the assumed distance,
our result confirms the validity of our earlier results about the stellar
populations in Tr~14 -- 16, NGC~3324, and NGC~3293 
\citep{CCCP-HAWKI,2011ApJS..194...11W,2011ApJS..194...12W,Preibisch14,2017A&A...605A..85P}.
The age estimates for the individual clusters in the CNC are
particularly important for reconstructing the star formation history
in the CNC. In this context, our finding that NGC~3293 is at the same
distance as the clusters Tr~14 -- 16 in the central Carina Nebula,
provides an important confirmation that  NGC~3293 actually is the oldest ($\simeq 10$~Myr)
of the larger clusters in the CNC.
The picture of the complex spatio-temporal progression
of the star-formation activity in the CNC outlined in 
\citet{2017A&A...605A..85P} is thus confirmed.

Our analysis of young foreground cluster vdBH~99, which overlaps with the
northwestern parts of the CNC, suggests an age of $\simeq 60$~Myr and 
a distance of $ (441 \pm 2)$~pc. Although this cluster comprises at least about
700 member stars, the contamination rate of the existing \textit{Chandra}
X-ray observations of the CCCP, NGC~3324, and NGC~3293 due to vdBH~99 members
is not more than a few percent.

\bigskip

\begin{acknowledgements}
This work has made use of data from the European Space Agency (ESA) mission
{\it Gaia} (\url{https://www.cosmos.esa.int/gaia}), processed by the {\it Gaia}
Data Processing and Analysis Consortium (DPAC,
\url{https://www.cosmos.esa.int/web/gaia/dpac/consortium}). Funding for the DPAC
has been provided by national institutions, in particular the institutions
participating in the {\it Gaia} Multilateral Agreement.
The research of T.P.~was partly supported by the Excellence Cluster ORIGINS 
which is funded by the 
Deutsche Forschungsgemeinschaft (DFG, German Research Foundation) under 
Germany's Excellence Strategy -  EXC-2094 - 390783311.
\end{acknowledgements}

\bibliographystyle{aa}
\bibliography{42576-ref}

\begin{appendix}

\section{Distance estimates with RUWE selection\label{RUWE.app}}

The RUWE values listed in the Gaia catalog
are a goodness-of-fit statistic
describing the quality of the astrometric solution
\citep[see][]{LL:LL-124}.
Values above 1.4 indicate a low reliability of the astrometric parameters
\citep{2021A&A...649A...5F}.
In this section, we investigate how our distance estimate would change
if we excluded all sources with RUWE $> 1.4$ from our samples.

Application of this RUWE selection criterion reduces the OB-star
samples of the Carina Nebula, NGC~3324, and NGC~3293 by 
20, 1, and 3 stars, respectively, 
leading to a 10.1\% reduction in the combined sample size.
The corresponding X-ray samples are reduced by 915, 45, and 54 objects,
leading to a 10.6\% reduction in the combined sample size.
We repeated all our distance calculations listed in Table \ref{table:distances}
with the RUWE selection criterion; the resulting distances are summarized in 
Table \ref{table:distancesRUWE}.
Comparison of the distances with and without the RUWE selection criterion
show only very small (and insignificant) differences 
for almost all subsamples.

For the Carina Nebula and the NGC~3293 OB star samples, the changes of the 
ML and also the \textit{Kalkayotl} distances are $\leq 2$~pc.
In the NGC~3324 OB star sample, which consists of only five stars,
the removal of the star HD~92\,206~C (which has RUWE = 5.27 and a
discrepant parallax qualifying it as a ``$2\sigma$'' outlier)
leads to a 8~pc change
in the ML mean distance and a 17~pc change in the \textit{Kalkayotl} distance
for this group when applying the $3\sigma$ outlier exclusion,
but no change for the $2\sigma$ and $1\sigma$ outlier exclusion case.
For the combined OB sample, the changes in the ML and the 
\textit{Kalkayotl} distances are $\leq 2$~pc.
Our final distance estimate for the OB stars of $2.36^{+0.05}_{-0.05}$~kpc
does not change.

For the X-ray samples,  the changes of the ML mean distances  
are always $\leq 5$~pc, and $\leq 2$~pc for the combined X-ray sample.
The \textit{Kalkayotl} distances for the X-ray samples change by $\leq 6$~pc,
and the changes for the combined X-ray sample are $\leq 3$~pc.
Our final distance estimate for the X-ray-selected stars of $2.34^{+0.05}_{-0.06}$~kpc 
does not change.
For the vdBH~99 sample, excluding the 27 stars with RUWE $> 1.4,$ this leads to
a \textit{Kalkayotl} distance of $440.7^{+2.3}_{-2.3}$~pc, which is only 0.1~pc
smaller than without RUWE selection.

\begin{table*}[hb]  
\caption{Maximum likelihood and \textit{Kalkayotl} estimates for the
distances for samples excluding stars with RUWE $> 1.4$.}
\label{table:distancesRUWE}
     \centering
\begin{tabular}{l|rrr|crc|rc}
\hline\hline                 
  Group &  \multicolumn{3}{c|}{outlier exclusion:} & ML parallax & 
\multicolumn{2}{c}{ML distance}  & 
\multicolumn{2}{|c}{\textit{Kalkayotl} distance}   \\
&  & {\small fore-} & {\small back-} & $\varpi_{\rm ML}$ & $D_{\rm ML}$ 
&  $2 \sigma$-range & $D_{\rm Kal}$ & {\small central 68.3\% quant.}\\
   &  & \multicolumn{2}{c|}{{\small ground}} & [mas] & 
\multicolumn{2}{c}{[kpc]} & \multicolumn{2}{|c}{[kpc]} \\
\hline                        
Carina Neb.& $3\sigma$ & 1 & 2 & $0.42457\pm 0.00156$ & $2.355$  &
$[ 2.338 \, , \, 2.373 ]$ & 2.366 & $[ 2.310 \, , \, 2.422 ]$\\
OB stars     & $2\sigma$ & 2 & 3 & $0.42457\pm 0.00158$ & $2.355$  &
$[ 2.338 \, , \, 2.373 ]$  & 2.367 & $[ 2.312 \, , \, 2.423]$ \\
  $(N = 113)$ & $1\sigma$ &  3 & 8 & $0.42567\pm 0.00160$ & $2.349$ & 
$[ 2.332 \, , \, 2.367 ]$  & 2.363 & $[2.308 \, , \, 2.418]$\\
\hline                        
  NGC~3324& $3\sigma$ &  0 & 0 & $0.43585\pm 0.00735$ & $2.294$ &
$[2.219 \, , \, 2.374]$  & 2.338 & $[2.267 \, , \, 2.410]$ \\
  OB stars & $2\sigma$ &  0 & 0 & $0.43585\pm 0.00735$ & $2.294$  &
$[2.219 \, , \, 2.374]$  & 2.338 & $[2.267 \, , \, 2.410]$ \\
  $(N = 4)$ & $1\sigma$ &  0 & 0 & $0.43585\pm 0.00735$ & $2.294$  &
$[2.219 \, , \, 2.374]$ &2.338 & $[2.267 \, , \, 2.410]$ \\
\hline                        
  NGC~3293& $3\sigma$ & 1 & 1 & $0.42539\pm 0.00178$ & $2.351$  & 
$[2.331 \, , \, 2.371]$ & 2.354 & $[ 2.298 \, , \, 2.410 ]$\\
  OB stars   & $2\sigma$ & 1 & 1 & $0.42539\pm 0.00178$ & $2.351$ & 
$[ 2.331 \, , \, 2.371 ]$ & 2.354 & $[ 2.298 \, , \, 2.410 ]$ \\
$(N = 96)$ & $1\sigma$ &  2 & 1 & $0.42471\pm 0.00179$ & $2.355$  & 
$[ 2.335 \, , \, 2.375 ]$ & 2.357 & $[ 2.302 \, , \, 2.412 ]$\\
\hline                        
All& $3\sigma$ & 2 & 3 & $0.42520\pm 0.00116$ & $2.352$  & 
$[2.339 \, , \, 2.365]$ & 2.362 & $[ 2.308 \, , \, 2.416 ]$\\
  OB stars   & $2\sigma$ & 3 & 4 & $0.42521\pm 0.00117$ & $2.352$  & 
$[2.339 \, , \, 2.365]$ & 2.363 & $[ 2.308 \, , \, 2.419 ]$ \\
$(N = 213)$ & $1\sigma$ &  5 & 9 & $0.42552\pm 0.00118$ & $2.350$  & 
$[2.337 \, , \, 2.363]$ & 2.362  & $[ 2.306 \, , \, 2.416 ]$\\
\hline\hline
CCCP& $3\sigma$ & 732 & 132 & $0.43052\pm 0.00066$ & $2.323$ 
& $[ 2.316 \, , \, 2.330]$ & 2.343 & $[2.290 \, , \, 2.397 ]$\\
X-ray      & $2\sigma$ & 934 & 418 & $0.42997\pm 0.00068$ & $2.326$ 
& $[ 2.318 \, , \, 2.333 ]$ & 2.339 & $[ 2.286 \, , \, 2.392 ]$\\
$(N = 7590)$ & $1\sigma$ &  1569 & 1280 & $0.43000\pm 0.00071$ & 
$2.326$  & $[ 2.318 \, , \, 2.333 ]$ & 2.331 & $[ 2.278 \, , \, 2.385 ]$\\
\hline                        
NGC~3324& $3\sigma$ & 52 & 9 & $0.42379\pm 0.00284$ & $2.360$  & 
$[2.328 \, , \, 2.392 ]$ & 2.345 & $[ 2.288 \, , \, 2.403 ]$\\
X-ray & $2\sigma$ & 65 & 19 & $0.42172\pm 0.00287$ & $2.371$  & 
$[2.339 \, , \, 2.404 ]$  & 2.358 & $[ 2.300 \, , \, 2.416 ]$\\
$(N = 410)$ & $1\sigma$ &  96 & 51 & $0.42379\pm 0.00306$ & $2.360$ 
& $[2.326 \, , \, 2.394 ]$ & 2.357 & $[ 2.299 \, , \, 2.414 ]$\\
\hline                        
NGC~3293& $3\sigma$ & 46 & 5 & $0.43160\pm 0.00221$ & $2.317$  & 
$[2.293 \, , \, 2.341 ]$ & 2.344 & $[ 2.289 \, , \, 2.399 ]$\\
X-ray & $2\sigma$ & 56 & 23 & $0.43219\pm 0.00222$ & $2.314$  & 
$[2.290 \, , \, 2.338 ]$ & 2.340 & $[ 2.284 \, , \, 2.395 ]$ \\
$(N = 548)$ & $1\sigma$ &  100 & 75 & $0.43130\pm 0.00227$ & $2.319$ 
& $[2.294 \, , \, 2.343 ]$ & 2.340 & $[ 2.284\, , \,2.396 ]$ \\
\hline                        
All& $3\sigma$ & 830 & 146 & $0.43028\pm 0.00062$ & $2.324$ 
  & $[2.317 \, , \, 2.331 ]$ & 2.341 & $[ 2.287 \, , \, 2.395 ]$\\
X-ray  & $2\sigma$ & 1055 & 460 & $0.42974\pm 0.00064$ & $2.327$ 
  & $[ 2.320 \, , \, 2.334 ]$ & 2.337 & $[ 2.284 \, , \, 2.391 ]$ \\
$(N = 8548)$ & $1\sigma$ &  1765 & 1406 & $0.42982\pm 0.00066$ & 
$2.327$  & $[ 2.319 \, , \, 2.334 ]$ & 2.331 & $[ 2.278 \, , \, 2.384 ]$ \\
\hline                                   
\end{tabular}
\end{table*}

\end{appendix}

\end{document}